\documentclass{PoS}

\newcommand{\im}{{\mbox{Im}\,}}

\newcommand{\ima}{\hbox{Im}\,}
\newcommand{\rea}{\hbox{Re}\,}

\usepackage{amsmath}
\usepackage{amsfonts}
\usepackage{amssymb}
\usepackage{graphicx}
\usepackage{dcolumn}
\usepackage{bm}

\title{Determination of Chiral Perturbarion Theory 
low energy constants from a precise 
description of $\pi\pi$ scattering threshold parameters}

\ShortTitle{Determination of ChPT LECs from a precise 
description of $\pi\pi$ scattering threshold parameters}

\author{\speaker{Guillermo Rios}\\
        Universidad de Murcia\\
        E-mail: \email{g.rios.marquez@um.es}}

\author{Jenifer Nebreda\\
        Yukawa Institute for Theoretical Physics, Kyoto University\\
        E-mail: \email{jnebreda@yukawa.kyoto-u.ac.jp}}

\author{Jose R. Pelaez\\
        Universidad Complutense de Madrid\\
        E-mail: \email{jrpelaez@fis.ucm.es}}

\abstract{We report on our 
determination of the values of the one and two loop low energy constants appearing
in the Chiral Perturbation Theory calculation of the $\pi\pi$ scattering
amplitude. For this
we use a  precise sum rule determination of scattering lengths and 
slopes that appear in the effective range expansion. In addition
we provide new sum rules and the values for these coefficients
up to third order in the expansion.
Our results when using only the scattering lengths and slopes of the S, P, D and F waves
are consistent with previous determinations, but seem to require 
higher order contributions if they are to 
accommodate the third order coefficients of the 
effective range expansion.}

\FullConference{The 7th International Workshop on Chiral Dynamics,\\
		August 6 -10, 2012\\
		Jefferson Lab, Newport News, Virginia, USA}

\begin{document}

\section{Introduction}
\vspace{-1mm}

Chiral Perturbation Theory (ChPT) \cite{chpt,Gasser:1983yg}
describes the low energy interactions of pions, which are the pseudo-Goldstone
bosons associated to the spontaneous chiral symmetry breaking of QCD, and hence,
the relevant degrees of freedom at low energies.
ChPT is built as the most general low energy expansion 
in terms of the pion momenta and mass that is compatible with the QCD symmetries. 
The details of the underlying dynamics at higher energies 
are encoded in a set of parameters, known
as low energy constants (LECs), that appear at the different orders 
in the expansion, and, once renormalized, absorb the loop divergences present at that order.
Since perturbative QCD cannot be applied 
at very low energies, it is particularly difficult to obtain the values 
of these LECs from first principles and, with few exceptions, 
the LECs have been determined best from the comparison with
experiment \cite{Gasser:1983yg,Riggenbach:1990zp,Amoros:2000mc,Colangelo:2001df}.
There are determinations of the LECs from lattice QCD 
(we refer to \cite{Colangelo:2010et} for a recent compilation), and 
it is also possible to obtain positivity constraints for the LECs coming from axiomatic
field theory \cite{axiomatic}.

In the ChPT  $\pi\pi$ scattering amplitude only certain
combinations of the LECs appear to a given order of the calculation.
To leading order, $O(p^2)$, there are no LECs and only the pion mass
decay constant $M_\pi$ and $f_\pi$ appear. To next to leading order (NLO), or $O(p^4)$,
which corresponds to a one-loop calculation,
only four LECs, called $l_1,...,l_4$ appear in the amplitude. At next to next to leading order (NNLO), or $O(p^6)$,
which corresponds to a two-loop calculation, only six independent terms
appear \cite{Knecht:1995tr}, multiplied by the corresponding combination of LECs,
denoted $\bar b_i$, $i=1...6$. 

Here we report our work \cite{original} on the determination of the LECs that
appear in the $\pi\pi$ scattering amplitude.
We obtain these LECs from fits to the coefficients of the momentum expansion of the
amplitude around threshold, known as threshold parameters, which can
be calculated within ChPT and expressed in terms of the LECs.
The experimental value of the threshold parameters we use is the one obtained from a 
dispersive data analysis in \cite{GarciaMartin:2011cn}, which includes the
very precise and reliable results on $K_{l4}$ decays from
the NA48/2 collaboration~\cite{Batley:2010zza}. We also provide here the
calculation of the third order coefficients of the threshold expansion, 
which was not performed in \cite{GarciaMartin:2011cn}.

\vspace{-1mm}
\section{Threshold parameters}
\vspace{-1mm}

The amplitude for  $\pi\pi$ scattering is customarily decomposed 
in terms of partial waves $t^{I}_\ell$, 
of definite isospin $I$ and angular momentum $\ell$:
$t^{I}_\ell(s)=\frac{1}{64\pi}\int_{-1}^1\, T^{I}(s,t,u) P_\ell(\cos\theta)\,d(\cos\theta),$
$\theta$ 
being the scattering angle, $P_\ell$ the Legendre polynomials, $s,t,u$ 
the usual Mandelstam variables
and $T$ the amplitude. 
With this normalization, the threshold expansion can be written as:
\begin{equation}
  \label{eq:effrange}
  \frac{1}{M_\pi} \rea t^{I}_\ell(s)= p^{2\ell}\Big( a_{\ell I}+b_{\ell I}\, p^2+\frac{1}{2}c_{\ell I} \,p^4+...\Big),
\end{equation}
where the $a_{\ell I}$ are usually called scattering lengths, the $b_{\ell I}$ slope parameters,
the $c_{\ell I}$ shape parameters,
and all of them, generically, threshold parameters.


The use of sum rules to obtain the values of threshold parameters is a well 
established technique \cite{Colangelo:2001df,thressumrules}
that we will also use here.
Our experimental determination of the threshold parameters
is performed using the 
parametrizations of \cite{GarciaMartin:2011cn}, which were obtained by highly 
constraining data fits to satisfy three sets of dispersion relations within 
uncertainties. In \cite{GarciaMartin:2011cn}, the values of the $a$ and $b$ 
parameters up to F waves were provided. With the aim of minimizing 
the uncertainties, they were obtained from sum rules,
 with the only exception of the $5 a_{S0}+2a_{S2}$ combination, which is orthogonal to the one appearing in the Olsson sum rule (note the spectroscopic notation, 
where the $\ell=0,1,2,3...$ are denoted S,P,D,F...). 

Here, we also provide
the calculation of the third order coefficients $c$ of the threshold expansion, 
which adds five more observables for the fit.
For the $c$ parameters with $\ell>0$, we use the Froissart-Gribov sum rules:
\begin{eqnarray}  \label{eq:FG}
\hspace{-4mm}c_{\ell I}=\frac{\sqrt{\pi}\,\Gamma(\ell+1)}{M_\pi\,\Gamma(\ell+3/2)}
\!\int_{4M_\pi^2}^{\infty} \!\!\!\!ds
&&\left\{\frac{16 \,{\im F^{I}}''(s,4M_\pi^2)}{(s-4M_\pi^2)^2 s^{\ell+1}} \right.\\
&&\hspace{3mm}\!-\!8(\ell+1)\frac{\im {F^{I}}'(s,4M_\pi^2)}{(s-4M_\pi^2)s^{\ell+2}}
\left.+\frac{\im F^{I}(s,4M_\pi^2)}{s^{\ell+3}} \frac{(\ell+2)^2(\ell+1)}{\ell+3/2}\right\},\notag
\end{eqnarray}
where $F^I(s,t)=T^I(s,t)/4\pi^2$ and the primes denote the derivative with respect to $\cos \theta$. 
This formula allow us to calculate the $c$ parameters for the P, D0, D2 and F waves.
For the S waves we provide two new sum rules, and also one for $c_P$, 
in order to reduce its error:
\begin{eqnarray}
\hspace{-4mm}c_{S2}&=&-6b_P-10a_{D2}+\frac{8}{M_\pi}\int_{4M_\pi^2}^\infty ds \left\{\frac{\ima F^{0+}(s,0)}{s^3}\right.\\
&+&\frac{1}{(s-4M_\pi^2)^{5/2}}\left.\left[\frac{\ima F^{0+}(s,0)}{\sqrt{s-4M_\pi^2}}
-\frac{2M_\pi a_{S2}^2}{\pi}-
\frac{s-4M_\pi^2}{\pi}
\left(\frac{M_\pi}{2}(2 a_{S2}b_{S2}+a_{S2}^4)-\frac{a_{S2}^2}{4M_\pi}\right)
\right]
\right\},\notag
\end{eqnarray}
\begin{eqnarray} 
&&\hspace{-4mm}c_{S0}=-2c_{S2}-20a_{D2}-10a_{D0}+\frac{12}{M_\pi}\int_{4M_\pi^2}^\infty ds \left\{\frac{\ima F^{00}(s,0)}{s^3}+\frac{1}{(s-4M_\pi^2)^{5/2}}\left[\frac{\ima F^{00}(s,0)}{\sqrt{s-4M_\pi^2}}\right.\right.\\
&&\hspace{.5cm}\left.\left.
-\frac{4M_\pi (2a_{S2}^2+a_{S0}^2)}{3\pi}-
\frac{s-4M_\pi^2}{3\pi}
\left(M_\pi[2 (2a_{S2}b_{S2}+a_{S2}^4)+2a_{S0}b_{S0}+a_{S0}^4]-\frac{2a_{S2}^2+a_{S0}^2}{2M_\pi}\right)
\right]
\right\}\nonumber
\end{eqnarray}
\begin{eqnarray}
\hspace{-6mm}c_P&=&-\frac{14 \,a_F}{3}+\frac{16}{3 M_\pi}
\int_{4M_\pi^2}^{\infty} ds
\left\{ \frac{\ima F^{I=0}(s,0)}{3s^4}-\frac{\ima F^{I=1}(s,0)}{2s^4}\right.\\
&&\hspace{4cm}\left.-\frac{5\ima F^{I=2}(s,0)}{6s^4} 
+\left[\frac{\ima F^{I=1}(s,0)}{(s-4M_\pi^2)^4}-\frac{3 a_P^2 M_\pi }{4\pi(s-4M_\pi^2)^{3/2}}\right]\right\}.\notag
\label{eq:newSRcp} 
\end{eqnarray}
 The derivation is similar to that of 
the sum rules for $b_P$, $b_{S0}$ and $b_{S2}$ obtained 
in \cite{sumrules}. They correspond to the threshold limit, taken from above,
 of the second derivative of a forward dispersion relation for the 
$F^{I=1}$, $F^{0+}$ and $F^{00}$ amplitudes, 
respectively. Let us recall that $F^{0+}=F^{I=2}/2+F^{I=1}/2$ 
whereas $F^{00}=2 F^{I=2}/3+F^{I=0}/3$. For a list of the resulting values, we refer the reader to Table II of our original work~\cite{original}. 

\vspace{-1mm}
\section{$O(p^4)$ fits}
\label{sec:op4fits}
\vspace{-1mm}

We start by fitting the $O(p^4)$ LECs by using the one-loop expression of the 
threshold parameters, which will help us check the stability of the LECs 
values and the need for higher orders. We actually fit the $\bar l_i$ 
parameters, which are basically the $l^r_i(\mu)$
at the $\mu=M_\pi$ scale and normalized so that they have values of order 
one~\cite{Gasser:1983yg}.
Note, however, that $\bar l_3$ and $\bar l_4$ only appear through the 
quark mass dependence of $M_\pi$ and $f_\pi$, respectively, and therefore we 
cannot expect much sensitivity to these two parameters from fits to the coefficients
of the momentum expansion of amplitudes. 
In addition, since the LECs only appear in the polynomial part of the partial waves, 
which at one loop is of $O(p^4)$, only ten observables carry any dependence on 
the LECs: $a_{S0)}$, $a_{S2)}$, $a_P$, $b_{S0}$, $b_{S2}$, $b_P$, $c_{S0}$, $c_{S2}$,
 $a_{D0}$ and $a_{D2}$. 
The rest of the coefficients multiply powers of the momentum higher than $p^4$ 
and thus, do not receive a contribution from the  $O(p^4)$ LECs.

In Table~\ref{fits-op4} we show the results of  our fits.
 First, we have fitted only the observables whose leading contribution is 
of $O(p^2)$, since these might be
more stable under the higher order corrections. The fit comes out with relatively low 
$\chi^2/d.o.f.$. Next we present two determinations of $\bar l_1$ and $\bar l_2$,
which can be fixed using only either $a_{D0}$ 
and $a_{D2}$, or $c_{S0}$ and $c_{S2}$. It is evident that the resulting values from 
those fits are incompatible. 
The incompatibility is even worse when fitting simultaneously 
the ten observables that depend on $\bar l_i$ to $O(p^4)$, where we 
obtain a high $\chi^2/d.o.f.$ value. 
Finally, the effect of higher order corrections has been studied by fitting
to the one-loop amplitude but replacing $f_\pi$ by $f_0$ in the $O(p^4)$ terms,
since the two expressions only differ in higher order contributions. This we show in row 5 of Table~\ref{fits-op4}.
The $\chi^2/d.o.f$ is somewhat lower, but the values of the LECs
come out rather different from the previous calculation. 
\begin{table*}[t]
{\small
 \centering
  \renewcommand{\arraystretch}{1.2}
  \begin{center}
    \begin{tabular}{lccccc}
      \hline \hline
      Fit to& $\bar l_1$ &	 $\bar l_2$ & $\bar l_3$ & $\bar l_4$ & $\chi^2/d.o.f.$ \\
      \hline
      $a_S, b_S, a_P$ & 1.1$\pm$1.0 &   5.1$\pm$0.7 &  $-1\pm$8 &   7.1$\pm$0.7 &
      0.23 \\
      $a_D$ & $-1.75\pm$0.22 & 5.91$\pm$0.10 & --- & --- & 0  \\
      $c_S$   & $-2.4\pm$0.9 &   4.8$\pm$0.4 & --- &  --- & 
      0  \\ 
      $a_S, b_S, a_P,a_D,c_S,b_P$ &  $-2.06\pm$0.14 &   5.97$\pm$0.07 &  $-5\pm$8 &   7.1$\pm$0.6 &
      7.9  \\ 
      $a_S, b_S, a_P,a_D,c_S,b_P$,  using $f_0$   &  $-1.06\pm$0.11 &   4.6$\pm$0.9 &  0$\pm$6 &   5.0$\pm$0.3 &
      7.06  \\ 
      \hline
      Estimate $O(p^4)$ & $-1.5\pm$0.5 & 5.3$\pm$0.7 & $-3\pm$7 & 6.0$\pm$1.2 & ---  \\ 
      \hline
    \end{tabular}
  \end{center}
  \caption{$O(p^4)$ fits to different sets of threshold parameters containing polynomial $O(p^4)$ contributions. We observe that a precise description of the observables is not possible at one loop. Anyway, we provide an estimate of how much one should enlarge the uncertainties of the LECs if, for simplicity, one still insists in using the one-loop formalism.}
  \label{fits-op4}
  }
\end{table*}

These results imply that, to the present level of precision, 
the one-loop ChPT formalism is not enough and calls for higher order corrections.
If one still wants to use this simpler version instead of the full two-loop amplitude
one can include the effect of higher orders into a systematic uncertainty of the LECs.
We propose to take the  weighted average of the two previous fits, 
including a systematic 
uncertainty to cover the LECs values of both fits. 
In Table IV of the original reference~\cite{original}, we compare 
the resulting threshold parameters obtained using this averaged set with 
the experimental values. Thanks to the larger uncertainty, 
the threshold parameters obtained are compatible within errors
with the experimental values, except for $b_{S0}$ and $b_P$, which differ by 
more than three and two standard deviations respectively. It is worth noting also that
the positivity constraints obtained from first principles \cite{axiomatic}
are perfectly satisfied by this set of LECs, even in the worst case scenario.

\vspace{-1mm}
\section{$O(p^6)$ fits}
\label{sec:op6fits}
\vspace{-1mm}

As commented in the introduction, the two-loop $\pi\pi$
scattering amplitude can be recast in terms of six independent terms
multiplied by their corresponding low energy constants $\bar b_i$.
In turn, these $\bar b_i$ can be rewritten in terms of the 
four $O(p^4)$ LECs and six combinations $r_i$ of $O(p^6)$ LECs \cite{Bijnens:1995yn}. 
The difference in the amplitude using one way or the other is $O(p^8)$.
However, despite increasing the number of parameters to ten, 
the $O(p^6)$ amplitude still provides just six independent structures.
As a consequence, the fits in terms of  $\bar l_i$ and $ r_i$ are
much more unstable, and can even lead to spurious solutions. 
For this reason we focus on the fits in terms of $\bar b_i$, 
and refer to the appendix~\cite{original} 
for a study of the $\bar l_i$, $r_i$ fits.

We first fit the ten threshold parameters used in the previous section because,
having a non-zero $O(p^4)$ polynomial contribution, we expect
these to be more stable under higher order corrections.
In the first row of 
Table~\ref{tab:fullbs} we show the resulting $\bar b_i$, 
which describes fairly well the fitted observables with a $\chi^2/d.o.f.= 1.2$. 
However, when fitting all 18 observables, we obtain somewhat different 
LECs (see the second row of Table~\ref{tab:fullbs}) and the $\chi^2/d.o.f.$ 
comes out rather poor. We have noticed that  $c_P$  
alone contributes almost to one third of the total $\chi^2$. 
This might indicate that $c_P$
receives important higher order contributions that are not being taken
into account in the $O(p^6)$ calculation.
Once again we obtain a crude estimate of the size of higher order ChPT 
corrections, by changing $f_\pi$ by $f_0$ in
the last term of the expansion. $c_P$ suffers indeed the
largest change, by almost $80\%$. 

Thus, we proceed to fit again all threshold parameters except
$c_P$. The result is shown in the third row of Table~\ref{tab:fullbs}.
The fit quality improves sizably, but we still get a high $\chi^2/d.o.f.=2.9$, which
indicates that the two-loop calculation may not be enough to describe
even the remaining threshold parameters with their current
level of precision.
Again, we see the effect of higher order corrections by
making a fit replacing $f_\pi$ by $f_0$ in the $O(p^6)$ terms. 
We show the results in the fourth row of Table~\ref{tab:fullbs}. 
Surprisingly, we now obtain a good $\chi^2/d.o.f.=1.0$ and all LECs are
less than two standard deviations away from those obtained by  fitting only the
threshold parameters with an $O(p^4)$ polynomial part. 
We conclude that, by excluding $c_P$, the two-loop fit can give an 
acceptable description of the rest of threshold parameters.
For this reason, we have once more
made a weighted average of the two fits 
(the one using $f_\pi$ and the one using $f_0$) adding systematic uncertainties to 
cover both sets. This we show in the fifth row of Table~\ref{tab:fullbs},
where we can see that they are also quite compatible with previous 
determinations in the literature \cite{Colangelo:2001df}. Also, this LECs
satisfy again the axiomatic constraints \cite{axiomatic}.
\begin{table*}[t]
{\small
  \centering
    \renewcommand{\arraystretch}{1.2}
  \begin{tabular*}{\textwidth}{@{\extracolsep{\fill}}lccccccc}
    \hline \hline
    Fit to& $\bar b_1$ & $\bar b_2$& $\bar b_3$& $\bar b_4$& $\bar b_5$& $\bar b_6$&
    \hspace{-2mm}$\frac{\chi^2}{d.o.f.}$ \\
    \hline
    $a_S, b_S, a_P,a_D,c_S,b_P$ & -14$\pm$4 &   14.6$\pm$1.2 &  -0.29$\pm$0.05 & 
    0.76$\pm$0.02 &   0.1$\pm$1.1 &   2.2$\pm$0.2 &     \hspace{-2mm}1.2 \\ 
    All & -2$\pm$3 &    14.2$\pm$1.0 &  -0.39$\pm$0.04 & 0.746$\pm$0.013 &
    3.1$\pm$0.3 &    2.58$\pm$0.12 & \hspace{-2mm}5.2   \\ 
    All but $c_P$  & -6$\pm$3 & 15.9$\pm$1.0 & -0.36$\pm$0.04 & 0.753$\pm$0.013 &
    2.2$\pm$0.4 &  2.44$\pm$0.12 &  \hspace{-2mm}2.9  \\
    All but $c_P$, using $f_0$ & -12$\pm$3 & 13.9$\pm$0.9 & -0.30$\pm$0.04 &  
    0.726$\pm$0.013 &  1.0$\pm$0.3 &  1.93$\pm$0.08 &  \hspace{-2mm}1.04  \\
    \hline
    Estimate $O(p^6)$& -10.5$\pm$5.1 & 14.5$\pm$1.8 & -0.31$\pm$0.06 & 0.73$\pm$0.02 & 1.3$\pm$1.0 &
    2.1$\pm$0.4 & \hspace{-2mm}--- \\  
    \hline
    Ref. \cite{Colangelo:2001df} & -12.4$\pm$1.6  & 11.8$\pm$0.6 & -0.33$\pm$0.07 & 0.74$\pm$0.01 & 3.6$\pm$0.4 & 2.35$\pm$0.02 & \hspace{-2mm}---\\
  \end{tabular*}
    \caption{$O(p^6)$ fits. In the first row we only fit to observables containing polynomial $O(p^4)$ contributions. Next we show the fit to all the threshold parameters obtained in this work. The quality is rather poor, but most of the disagreement is caused by $c_P$. When this observable is omitted, the resulting fits are much better, specially when using $f_0$ instead of $f_\pi$ in the last term of the ChPT expansion. We provide an estimate of the LECs uncertainties from the fits to all observables except $c_P$, as a weighted average of the fits using $f_0$ or $f_\pi$. The resulting $\bar b_i$ parameters are very consistent with previous determinations, listed in the last row. 
  }
    \label{tab:fullbs}
}
  \end{table*}

\vspace{-1mm}  
\section{Summary}
\label{sec:conclusions}
\vspace{-1mm}

We have reported a work~\cite{original} on the determination the low energy constants of $SU(2)$ Chiral Perturbation Theory (ChPT) at one and two loops by fitting to the threshold parameters obtained from sum rules using a recent and precise dispersive analysis of data~\cite{GarciaMartin:2011cn}, together with six additional observables that we have studied here. 
  
We have checked that the one-loop formalism is clearly insufficient to accommodate the present level of precision. The $\chi^2/d.o.f.$ improves remarkably when using the two-loop expansion, although it is still not sufficient to get a good quality fit. This suggests that even higher order ChPT contributions may still be required to describe all these observables simultaneously.

\end{document}